\documentclass[11pt,a4paper]{article}

\usepackage{amsmath,amsfonts,amssymb}
\usepackage{algorithmic}
\usepackage{algorithm}
\usepackage{array}
\usepackage[caption=false,font=normalsize,labelfont=sf,textfont=sf]{subfig}
\usepackage{textcomp}
\usepackage{stfloats}
\usepackage{url}
\usepackage{graphicx}
\usepackage{cite}
\usepackage{booktabs}
\usepackage{multirow}
\usepackage{color}
\usepackage{bm}
\begin{document}

\title{A Conjugate Gradient Unrolled Network with PSF Conditioning for Non-Diagonal Data Fidelity in CASSI Reconstruction}

\author{Xiang~Li,~Qiuyu~Yue,~Yang~Zhang,~and~Zhenrong~Zheng\textsuperscript{*}
\thanks{\textsuperscript{*}Corresponding author: Zhenrong~Zheng (e-mail: zzr@zju.edu.cn).}
\thanks{All authors are with the College of Optical Science and Engineering, 
Zhejiang University, Hangzhou 310027, China.}}

\maketitle
\begin{abstract}
Deep unfolding methods achieve state-of-the-art performance in coded aperture snapshot spectral imaging (CASSI) reconstruction but typically rely on closed-form data-fidelity updates that assume an ideal optical system. In practical CASSI systems, the field-dependent and wavelength-dependent point spread function (PSF) introduces convolutional coupling that breaks the diagonal structure of the normal equation, rendering closed-form updates inapplicable. We propose a deep unfolding framework that addresses this fundamental algorithmic challenge through three contributions: (1)~a $K$-step conjugate gradient (CG) unrolling that explicitly solves the non-diagonal normal equation under PSF-inclusive forward operators; (2)~a learned gradient refinement module with wavelength-adaptive step sizes generated from a per-wavelength PSF embedding; and (3)~a PSF-conditioned penalty estimator that adapts the ADMM regularization strength to the optical degradation severity. A Monte Carlo PSF training strategy further improves robustness to manufacturing-induced PSF variations. Our method achieves 30.53~dB on the KAIST dataset, +2.73~dB over the DPU baseline (1.27M) using a comparable number of parameters (1.42M), and +1.70~dB over a larger baseline (DPU-B+, 2.12M) using 33\% fewer parameters.
\end{abstract}

\section{Introduction}
Coded aperture snapshot spectral imaging (CASSI) has emerged as a promising paradigm for acquiring three-dimensional hyperspectral images (HSIs) from a single two-dimensional compressed measurement~\cite{Gehm2007,Wagadarikar2009}. By leveraging a coded aperture mask and dispersive optics, CASSI systems encode the spatial-spectral information of a scene into a snapshot measurement, enabling high-throughput data acquisition without the need for temporal scanning. The reconstruction of the original HSI from this compressed measurement constitutes a highly ill-posed inverse problem, which has attracted significant research attention in recent years~\cite{MST2022,HDNet2022,DAUHST2022,RDLUF2023,DPU2024}.

Deep unfolding methods have demonstrated remarkable success in CASSI reconstruction by combining the interpretability of optimization-based algorithms with the representational power of deep neural          networks~\cite{DGSMP2021,HerosNet2022,DAUHST2022,RDLUF2023,GAPNet2023,DPU2024}. These methods formulate the reconstruction as an optimization problem and unroll the iterative solver into a fixed number of network stages, where each stage corresponds to one iteration with learnable components. Notable examples include ADMM-Net~\cite{ADMMNet2016}, DGSMP~\cite{DGSMP2021}, DAUHST~\cite{DAUHST2022}, RDLUF~\cite{RDLUF2023}, and the recent Dual Prior Unfolding (DPU)~\cite{DPU2024}, which achieves state-of-the-art performance by jointly exploiting image priors and degradation-associated priors within an augmented Lagrangian framework. However, these methods universally adopt a mask-only forward model $\mathbf{g} = \mathbf{\Phi}\mathbf{f} + \boldsymbol{\varepsilon}$, implicitly assuming that the optical point spread function (PSF) is ideal (i.e., a Dirac delta function).

In practice, every optical system introduces spatially-varying and wavelength-dependent PSF degradation due to residual aberrations. In CASSI systems, optical aberrations such as coma, astigmatism, and field curvature cause the PSF to vary across the field of view and with wavelength. The mismatch between the assumed mask-only forward model and the actual PSF-degraded imaging process introduces systematic reconstruction errors that cannot be compensated by increasing network capacity alone.

The importance of PSF modeling in CASSI has been recognized in prior work along two distinct lines. On the physical modeling side, Song~et~al.~\cite{Song2022} proposed a high-accuracy image formation model that explicitly incorporates the system PSF, estimated from calibration images via regularized least-squares, into the measurement matrix for traditional optimization-based reconstruction (e.g., total variation and non-local low-rank methods). Their work demonstrated that faithful PSF modeling significantly improves reconstruction quality over the simplified binary model. On the data-driven side, Yue~et~al.~\cite{Yue2024} computed the Huygens PSF from optical design parameters and used it both to generate realistic aberrated training data and as a conditional input to a cGAN-based reconstruction network, achieving improved robustness to optical aberrations without requiring physical calibration.

Despite these advances, a fundamental gap remains: \emph{no existing method integrates a PSF-inclusive forward operator into a deep unfolding reconstruction framework}. Song~et~al.'s approach~\cite{Song2022} pairs PSF-aware physics with traditional solvers that lack the representational power of learned priors. Yue~et~al.'s approach~\cite{Yue2024} leverages PSF information within a data-driven network but does not embed the PSF into an iterative physics-consistent optimization loop. Meanwhile, state-of-the-art deep unfolding methods achieve strong reconstruction performance but remain entirely PSF-agnostic. Bridging PSF-aware optical modeling with deep unfolding-based reconstruction is the central goal of this work.

The integration of PSF into a deep unfolding framework is not merely an engineering extension---it introduces a fundamental algorithmic challenge. In conventional unfolding methods such as DPU~\cite{DPU2024}, the forward model only accounts for coded aperture modulation and dispersive shifting, resulting in a diagonal normal matrix $\mathbf{\Phi}^T\mathbf{\Phi}$ that admits a closed-form solution for the data-fidelity subproblem. However, when PSF information is incorporated into the measurement matrix, this diagonal structure is broken and no closed-form solution exists, necessitating a new solver for the data-fidelity subproblem.

To address these challenges, we propose a PSF-aware deep unfolding network for CASSI reconstruction that integrates the PSF-inclusive forward operator $[\mathbf{A}(\mathbf{f})]_c = H_c * (\Phi_c \odot f_c)$, following the physical imaging model of Song~et~al.~\cite{Song2022}, into the iterative optimization structure of the DPU framework~\cite{DPU2024}. Since PSF inclusion renders the normal matrix $\mathbf{A}^T\mathbf{A}$ non-diagonal, we introduce a differentiable conjugate gradient (CG) solver with $K=2$ warm-started iterations. PSF information is further injected at the optimization level through two complementary pathways: (i)~a global PSF embedding conditions the ADMM penalty parameter, and (ii)~a per-wavelength PSF embedding generates wavelength-adaptive step sizes and gradient biases in the post-CG refinement step. A Monte Carlo PSF training strategy based on Zernike polynomial tolerance analysis exposes the network to realistic manufacturing-induced PSF variations.

We evaluate our method on the CAVE~\cite{CAVE2007} and KAIST~\cite{KAIST2017} datasets under field-dependent PSF degradation simulated from a real DD-CASSI optical design using Zemax. Simulation experiments show a 1.70~dB PSNR improvement over the larger PSF-agnostic baseline (DPU-B+) under nominal PSF conditions and 1.83~dB under Monte Carlo PSF conditions. And larger improvements are observed over other state-of-the-art unfolding methods (e.g., +3.14~dB over RDLUF, +2.39~dB over MST).

The main contributions of this work are summarized as follows:
\begin{itemize}
\item We propose, to the best of our knowledge, the first PSF-aware deep unfolding framework for CASSI, integrating field-dependent, wavelength-dependent PSF into the physical operators of each unfolding stage. Unlike prior work that pairs PSF modeling with traditional solvers~\cite{Song2022} or uses PSF as side information for data-driven networks~\cite{Yue2024}, our method embeds the PSF directly into the iterative optimization structure.

\item We introduce a $K$-step CG-unrolled data-fidelity solver as a necessary algorithmic consequence of the PSF-inclusive forward model, replacing the closed-form solution that becomes inapplicable when $\mathbf{A}^T\mathbf{A}$ is non-diagonal. The CG solver is fully differentiable and supports end-to-end training.

\item We propose a wavelength-adaptive gradient refinement module that uses a learned per-wavelength PSF embedding to generate spectral-channel-specific step sizes and gradient biases for the post-CG update, enabling the network to apply per-channel adaptive convergence rates based on PSF severity.

\item We propose a Monte Carlo PSF training strategy based on Zernike polynomial tolerance analysis, exposing the network to realistic manufacturing-induced PSF variations and improving robustness to off-design PSFs.
\end{itemize}

\section{Related Work}

\subsection{Snapshot Compressive Imaging Reconstruction}

The reconstruction of HSIs from CASSI measurements has been extensively studied using both model-based and learning-based approaches. Early model-based methods employ hand-crafted priors such as sparsity~\cite{Tan2016AMP}, low-rank structure~\cite{Liu2019Rank}, and total variation~\cite{Yuan2016GAPTV} within iterative optimization frameworks. While these methods offer interpretability, they are computationally expensive and often produce unsatisfactory results due to the limited expressiveness of hand-crafted priors.

Learning-based methods leverage deep neural networks to learn the mapping from compressed measurements to reconstructed HSIs. End-to-end approaches such as TSA-Net~\cite{TSANet2020}, HDNet~\cite{HDNet2022}, and $\lambda$-Net~\cite{LambdaNet2019} directly learn the reconstruction mapping without explicit optimization structure. Transformer-based methods such as MST~\cite{MST2022} and CST~\cite{CST2022} leverage self-attention mechanisms to capture non-local spatial-spectral correlations. While these methods achieve faster inference, they lack the interpretability and physical consistency of optimization-based approaches.

Deep unfolding methods bridge the gap between model-based and learning-based approaches by unrolling iterative optimization algorithms into trainable network architectures. DGSMP~\cite{DGSMP2021} employs a Gaussian scale mixture prior within a MAP estimation framework. DAUHST~\cite{DAUHST2022} introduces degradation-aware modules and half-shuffle attention to address the ill-posedness of the reconstruction. RDLUF~\cite{RDLUF2023} jointly exploits spatial and spectral priors with degradation learning. DPU~\cite{DPU2024} proposes a dual prior framework that simultaneously utilizes image priors and degradation-associated priors, achieving state-of-the-art performance with minimal computational cost.

Despite the diversity of these approaches, all existing methods assume an ideal optical system without PSF degradation. Our work extends the deep unfolding paradigm by incorporating a physically accurate PSF-inclusive forward model, addressing a fundamental gap in the current literature.

\subsection{PSF-Aware and Degradation-Aware Image Reconstruction}

The incorporation of degradation information into neural network-based image reconstruction has been explored in several contexts. In the image deblurring literature, non-blind deconvolution methods assume a known PSF and solve the inverse problem using Wiener filtering~\cite{Wiener1949}, Richardson-Lucy iteration~\cite{Richardson1972}, or learned approaches~\cite{Zhang2017IRCNN,Zamir2022Restormer}. Recent deep learning methods for spatially-varying deblurring include MIMO-UNet~\cite{Cho2021MIMOUNet} and Restormer~\cite{Zamir2022Restormer}, which implicitly handle spatially-varying degradation through large receptive fields but do not explicitly model the PSF.

In computational imaging, PSF-aware reconstruction has been explored for microscopy~\cite{Weigert2018ContentAware}, astronomical imaging~\cite{Hirsch2010Efficient}, and lensless imaging~\cite{Sinha2017Lensless}. These methods typically incorporate the PSF into the forward model and solve the resulting inverse problem using iterative algorithms or learned priors.

Within the CASSI literature, Song~et~al.~\cite{Song2022} proposed a high-accuracy image formation model that incorporates the system PSF into the measurement matrix, demonstrating significant reconstruction improvements with traditional optimization solvers (e.g., TV and non-local low-rank). Yue~et~al.~\cite{Yue2024} computed the Huygens PSF from optical design parameters to generate aberrated training data for a cGAN-based network, improving robustness without physical calibration. However, the former pairs PSF-aware physics with traditional solvers lacking learned priors, while the latter does not embed PSF into an iterative optimization structure.

Degradation-aware deep unfolding methods such as DAUHST~\cite{DAUHST2022} and RDLUF~\cite{RDLUF2023} consider the degradation induced by the coded aperture mask, including shift and compression effects. However, these methods do not account for the optical PSF, which represents a fundamentally different type of degradation that is spatially-varying, wavelength-dependent, and determined by the optical design rather than the coding pattern. Our work bridges this gap by integrating PSF-aware physical modeling into a deep unfolding framework, combining the interpretability of optimization-based methods with the representational power of learned priors.

\subsection{Deep Unfolding with Learned Physical Operators}

Deep unfolding methods traditionally employ fixed physical operators, including the forward model and its adjoint, within each iteration, with only the proximal operators or regularization terms being learned~\cite{ADMMNet2019,ISTANet2018}. Recent work has explored making the physical operators themselves adaptive. ADMM-Net~\cite{ADMMNet2016} learns penalty parameters across stages. Deep Tensor ADMM-Net~\cite{ADMMNet2019} extends this idea to tensor-structured problems. HeroSNet~\cite{HerosNet2022} improves inter-stage interaction and parameter adaptation.

Our work differs from these approaches in that we do not learn the physical operators themselves: the forward and adjoint operators are determined by the known PSF and mask. Instead, we address the computational challenge that arises when the physical operator includes the PSF: the normal equation in the f-subproblem becomes non-diagonal, requiring iterative solvers. Our CG unrolling approach provides a principled solution to this challenge while maintaining full differentiability for end-to-end training.

\section{Problem Formulation}

\subsection{Conventional CASSI Forward Model}
The dual-disperser coded aperture snapshot spectral imaging (DD-CASSI) system captures a three-dimensional hyperspectral image $\mathbf{F} \in \mathbb{R}^{H \times W \times \Lambda}$ into a single two-dimensional measurement $\mathbf{G} \in \mathbb{R}^{H \times W}$. In DD-CASSI, the first dispersive prism introduces wavelength-dependent spatial shifts, the coded aperture modulates the dispersed light, and the second prism reverses the dispersion before the detector integrates across the spectral dimension. Due to the dispersion, each spectral channel corresponds to a different region of the physical coded aperture $M \in \mathbb{R}^{H \times (W+(\Lambda-1)d)}$, where $d$ is the pixel shift per channel. Let $\Phi_i \in \mathbb{R}^{H \times W}$ denote the equivalent mask for the $i$-th channel, extracted from the corresponding region of $M$. The conventional forward model is
\begin{equation}
\label{eq:conv_forward}
G(m,n) = \sum_{i=1}^{\Lambda} \Phi_i(m,n) \odot F_i(m,n) + \epsilon(m,n),
\end{equation}
where $\odot$ denotes element-wise multiplication and $\epsilon$ is additive noise. In matrix-vector form:
\begin{equation}
\label{eq:conv_matrix}
\mathbf{g} = \mathbf{\Phi}\mathbf{f} + \boldsymbol{\epsilon}.
\end{equation}

Existing deep unfolding methods~\cite{DGSMP2021,DAUHST2022,RDLUF2023,DPU2024} adopt this PSF-free forward model, where $\mathbf{\Phi}^{T}\mathbf{\Phi}$ is diagonal and the data fidelity subproblem admits a closed-form solution.

\begin{figure*}[!t]
\centering
\includegraphics[width=\textwidth]{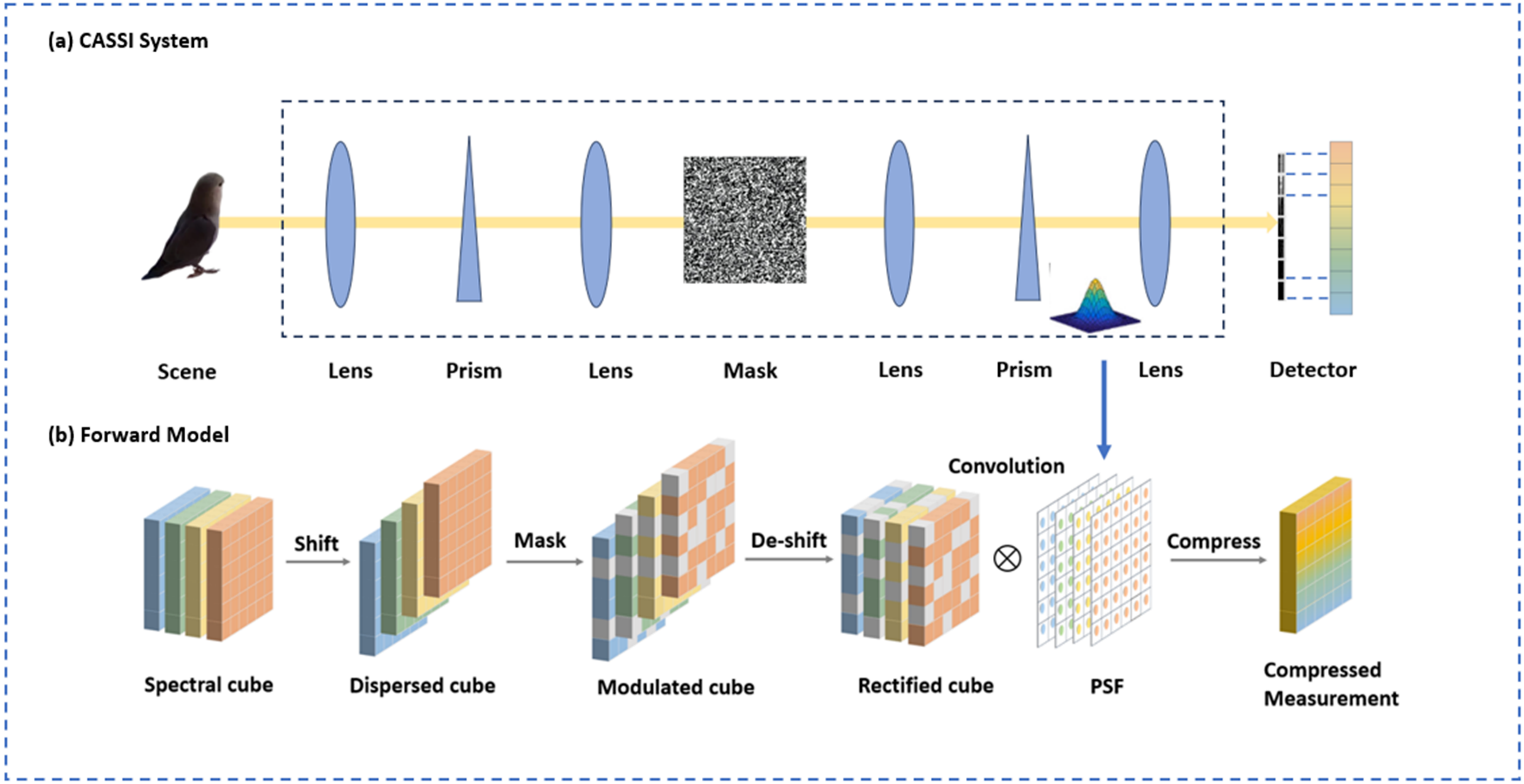}
\caption{The DD-CASSI imaging pipeline. (a) Physical optical layout consisting of two dispersive prisms, relay lenses, a coded aperture mask, and a detector. The spatially-varying, wavelength-dependent PSF is introduced by the imaging optics (indicated by the blue arrow). (b) Corresponding forward model: the input spectral cube undergoes spectral dispersion (shift), coded aperture modulation (mask), de-dispersion (de-shift), convolution with the system PSF, and spectral compression to produce the final 2D measurement. The PSF convolution step, typically ignored in existing reconstruction methods, is explicitly incorporated in our forward model.}
\label{fig:cassi_pipeline}
\end{figure*}

In practice, the imaging optics of a CASSI system introduce wavelength-dependent and spatially-varying point spread function (PSF) degradation due to residual optical aberrations. As shown in Fig.~\ref{fig:framework}, the actual imaging process includes an additional PSF blurring step that is absent from the conventional model. Following the physical imaging model proposed by Song~et~al.~\cite{Song2022}, we explicitly model this degradation by defining the PSF-inclusive forward operator as
\begin{equation}
\label{eq:psf_forward}
G(m,n) = \sum_{i=1}^{\Lambda} H_i * (\Phi_i(m,n) \odot F_i(m,n)) + \epsilon(m,n),
\end{equation}
The kernel $H_i$ represents the effective end-to-end PSF of
the DD-CASSI system at wavelength $\lambda_i$, obtained from
Zemax simulation of the complete optical path. Consistent
with the model in Song~et~al.~\cite{Song2022}, this PSF is applied after the coded aperture modulation
in~\eqref{eq:psf_forward}, which is a well-established
simplification in CASSI image formation.

In operator notation, the PSF-inclusive forward model becomes
\begin{equation}
\label{eq:psf_matrix}
\mathbf{g} = \mathbf{A}\mathbf{f} + \boldsymbol{\epsilon},
\end{equation}
where the forward operator $\mathbf{A}$ is defined channel-wise as
\begin{equation}
\label{eq:forward_op}
[\mathbf{A}(\mathbf{f})]_c = H_c * (\Phi_c \odot f_c).
\end{equation}
Note that within each reconstruction block, the PSF $H_c$ is assumed 
spatially invariant, while different blocks are assigned different 
field-dependent PSFs. This piecewise treatment approximates the 
continuous spatial variation of the true optical PSF at a spatial 
resolution determined by the field grid density.

The corresponding adjoint operator is
\begin{equation}
\label{eq:adjoint_op}
[\mathbf{A}^T(\mathbf{r})]_c = \Phi_c \odot (H_c^* * \mathbf{r}),
\end{equation}

where $H_c^*$ denotes the conjugate of $H_c$, which corresponds to complex conjugation in the Fourier domain and spatial flipping in the spatial domain. Both $\mathbf{A}$ and $\mathbf{A}^T$ are implemented efficiently via FFT-based convolution.

The inclusion of the PSF fundamentally changes the structure of the inverse problem. In the conventional model, $\mathbf{\Phi}^T\mathbf{\Phi}$ is diagonal, enabling closed-form inversion. In the PSF-inclusive model, $\mathbf{A}^T\mathbf{A}$ involves convolutional coupling through the PSF, making it non-diagonal and precluding the closed-form solution used in existing unfolding methods. This structural difference motivates the algorithmic modifications described in Section~IV.

\subsection{PSF Characterization via Zernike Polynomials}

The field-dependent PSF is characterized using Zernike polynomial coefficients obtained from optical design software, i.e., Zemax OpticStudio. For a DD-CASSI system with $N_f$ field positions and $\Lambda$ wavelengths, the Zernike coefficients $\{c_{k,\lambda,j}\}$ are exported for each field $k$, wavelength $\lambda$, and Zernike term $j$. In this work, terms $j=4,\ldots,15$ are used, corresponding to defocus through secondary spherical aberration; piston, tip, and tilt are excluded since they do not affect image sharpness.

Given the Zernike coefficients, the PSF for each field-wavelength combination is computed through a physical optics model:
\begin{equation}
\label{eq:zernike_psf}
H_{k,\lambda}(x,y) = \left| \mathcal{F}^{-1} \!\left\{ P(u,v) \cdot e^{i \frac{2\pi}{\lambda} \sum_{j} c_{k,\lambda,j} Z_j(u,v)} \right\} \right|^2
\end{equation}

where $P(u,v)$ is the circular pupil function, $Z_j(u,v)$ are the Zernike basis functions, and $\mathcal{F}^{-1}$ denotes the inverse Fourier transform. This physical layer is implemented as a fixed non-learnable module that converts Zernike coefficients to PSF kernels, as shown in the upper portion of Fig.~\ref{fig:framework}.

To account for manufacturing tolerances, we perform Monte Carlo (MC) tolerance analysis in Zemax and generate $N_{\mathrm{MC}}=1000$ realizations of the Zernike coefficients. Each realization represents a plausible manufactured lens with perturbed surface parameters. The resulting MC PSF distribution is used for data augmentation during training, as described in Section~\ref{sec:mc_training}.

\section{Proposed Method}
\label{sec:proposed_method}

\subsection{Overview}

\begin{figure*}[!t]
\centering
\includegraphics[width=\textwidth]{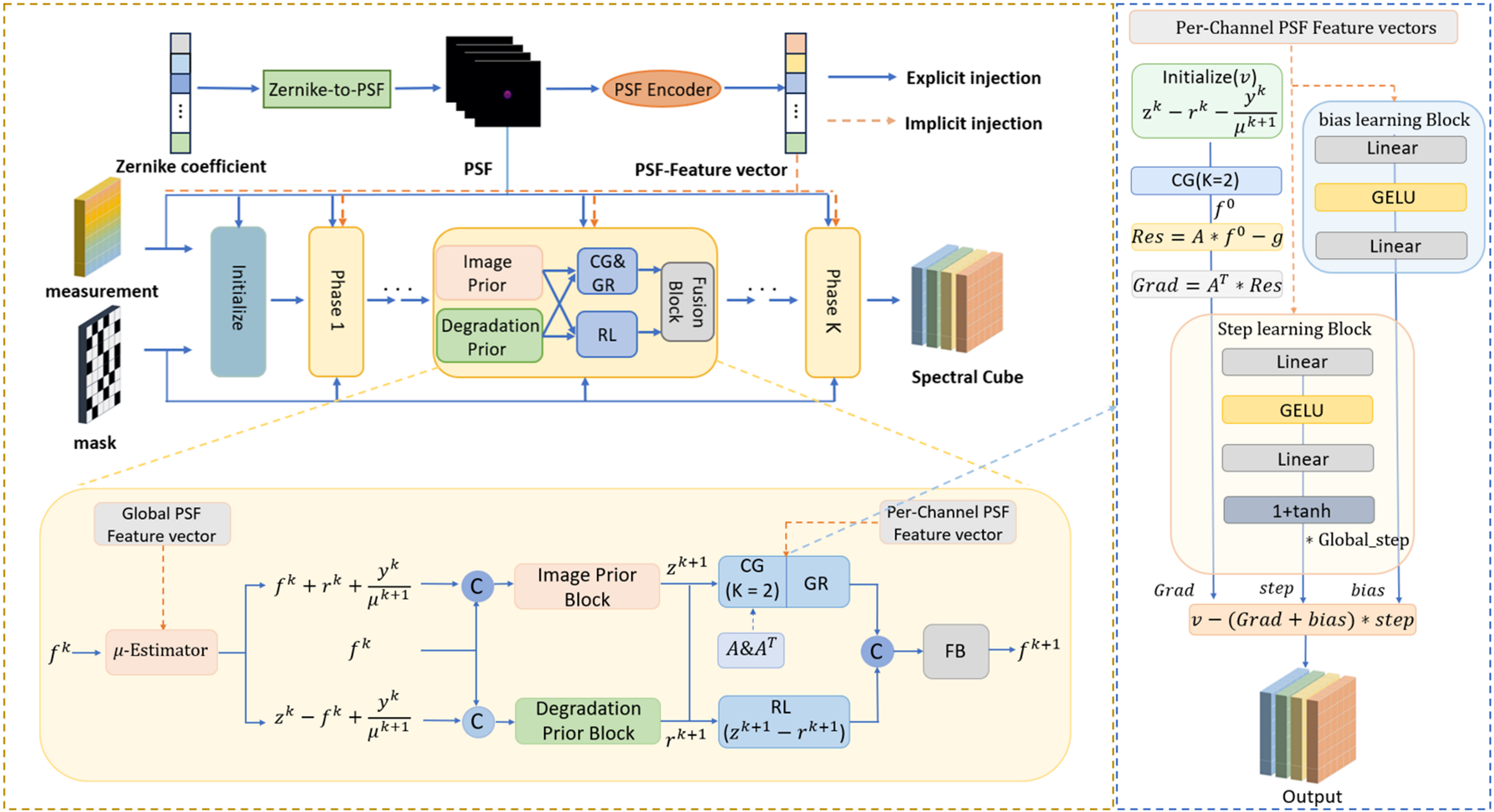}
\caption{Overall architecture of the proposed PSF-aware deep unfolding network. Top: Zernike coefficients are converted to field-dependent and wavelength-dependent PSFs through a fixed physical layer, then encoded by the PSF Encoder into a global feature vector and per-channel feature vectors. The reconstruction network receives PSF information through three pathways: explicit use of $\mathbf{A}$ and $\mathbf{A}^T$ in the CG solver (solid blue arrows), and embedding-based conditioning of the $\mu$ Estimator and gradient refinement module (dashed orange arrows). Middle and bottom-left: the ADMM-based reconstruction proceeds through $S$ unfolding stages. Within each stage, PSF embeddings are injected into the $\mu$ Estimator and the CG\&GR module, while the Image Prior Block, Degradation Prior Block, and Fusion Block focus on spatial-spectral feature learning. Right: internal structure of the CG and gradient refinement (GR) module. After $K=2$ CG iterations, the per-channel PSF features generate per-wavelength step scales (via the Step learning Block) and gradient biases (via the Bias learning Block), enabling wavelength-adaptive gradient updates.}
\label{fig:framework}
\end{figure*}

Our method builds upon the Dual Prior Unfolding (DPU) framework~\cite{DPU2024} and extends it with PSF-aware reconstruction capability. The overall architecture is illustrated in Fig.~\ref{fig:framework}. The network takes three inputs: the compressed measurement $\mathbf{g}$, the coded aperture mask $\mathbf{\Phi}$, and the field-dependent PSF $\mathbf{H}$. The PSF is first generated from Zernike coefficients through a fixed physical layer, and then encoded into a compact feature vector by a learnable PSF encoder. The ADMM-based reconstruction proceeds through $S$ unfolding stages, where PSF information is injected through three complementary pathways operating in the optimization space: (i)~explicit use of $\mathbf{A}$ and $\mathbf{A}^T$ in the CG solver, (ii)~PSF-conditioned penalty parameter estimation, and (iii)~wavelength-adaptive step generation in the gradient refinement module.

\subsection{ADMM Formulation with PSF-Inclusive Forward Model}

Following DPU~\cite{DPU2024}, we formulate the reconstruction as a constrained optimization problem with dual priors:
\begin{equation}
\label{eq:opt_problem}
\begin{aligned}
\arg\min_{\mathbf{f},\mathbf{z},\mathbf{r}} \quad &
\frac{1}{2}\|\mathbf{g}-\mathbf{A}\mathbf{f}\|_2^2
+ \gamma \mathcal{D}(\mathbf{z})
+ \tau \mathcal{R}(\mathbf{r}) \\
\text{s.t.} \quad &
\mathbf{f} = \mathbf{z} - \mathbf{r},
\end{aligned}
\end{equation}
where $\mathbf{A}$ is the PSF-inclusive forward operator defined in \eqref{eq:forward_op}, $\mathcal{D}(\cdot)$ is the image prior, $\mathcal{R}(\cdot)$ is the degradation prior, and $\gamma,\tau$ are tradeoff parameters.

Applying the Augmented Lagrangian Method (ALM), we obtain
\begin{equation}
\label{eq:alm}
\begin{aligned}
\mathcal{L}(\mathbf{f},\mathbf{z},\mathbf{r},\mathbf{y},\mu)
=&\frac{1}{2}\|\mathbf{g}-\mathbf{A}\mathbf{f}\|_2^2 \\
&+ \frac{\mu}{2}
\left\|
\mathbf{f}-\mathbf{z}+\mathbf{r}+\frac{\mathbf{y}}{\mu}
\right\|_2^2 \\
&+ \gamma \mathcal{D}(\mathbf{z})
+ \tau \mathcal{R}(\mathbf{r}),
\end{aligned}
\end{equation}
where $\mathbf{y}$ is the Lagrange multiplier and $\mu$ is the penalty parameter. The optimization proceeds by alternately updating $\mathbf{z}$, $\mathbf{r}$, $\mathbf{f}$, and $\mathbf{y}$ at each stage $k$. The detailed single-stage structure is shown in Fig.~\ref{fig:framework} (bottom)

\subsubsection{z-subproblem: Image Prior}

The preliminary restored image is obtained via the Image Prior Block (IPB):
\begin{equation}
\label{eq:z_update}
\mathbf{z}^{k+1}
=
\mathrm{IPB}
\left(
\left[
\mathbf{f}^k+\mathbf{r}^k+\frac{\mathbf{y}^k}{\mu^{k+1}},\;
\mathbf{f}^k
\right]
\right),
\end{equation}
where $[\cdot,\cdot]$ denotes channel-wise concatenation. The IPB follows the same asymmetric U-Net architecture as the original IPB in DPU~\cite{DPU2024} and does \emph{not} receive PSF information directly; PSF conditioning enters the network only through the CG solver, the gradient refinement step, and the penalty parameter (Section~\ref{sec:psf_conditioning}).

\subsubsection{r-subproblem: Degradation Prior}

The degradation residual is estimated via the Degradation Prior Block (DPB):
\begin{equation}
\label{eq:r_update}
\mathbf{r}^{k+1}
=
\mathrm{DPB}
\left(
\left[
\mathbf{z}^k-\mathbf{f}^k-\frac{\mathbf{y}^k}{\mu^{k+1}},\;
\mathbf{f}^k
\right],\;
\mathbf{\Phi},\;
\mathbf{A}(\mathbf{\Phi})
\right),
\end{equation}
where $\mathbf{A}(\mathbf{\Phi})$ denotes the PSF-convolved mask, providing the degradation prior with explicit information about how the mask pattern is degraded by the optical PSF. This is a key difference from the original DPB in DPU, which only uses the mask and its shifted/compressed version.

\subsubsection{f-subproblem: Data Fidelity}

Let
\begin{equation}
\label{eq:v_def}
\mathbf{v}
=
\mathbf{z}^{k+1}
-
\mathbf{r}^{k+1}
-
\frac{\mathbf{y}^{k}}{\mu^{k+1}}.
\end{equation}
The f-update requires solving
\begin{equation}
\label{eq:f_subproblem}
\mathbf{f}^{k+1}
=
\arg\min_{\mathbf{f}}
\|\mathbf{g}-\mathbf{A}\mathbf{f}\|_2^2
+
\mu^{k+1}\|\mathbf{f}-\mathbf{v}\|_2^2.
\end{equation}
Setting the gradient to zero yields the normal equation:
\begin{equation}
\label{eq:normal_eq}
(\mathbf{A}^{T}\mathbf{A}+\mu\mathbf{I})\mathbf{f}
=
\mathbf{A}^{T}\mathbf{g}
+
\mu\mathbf{v}
\triangleq
\mathbf{b}.
\end{equation}

In the original DPU~\cite{DPU2024}, where $\mathbf{A}=\mathbf{\Phi}$, the matrix $\mathbf{\Phi}^{T}\mathbf{\Phi}$ is diagonal and \eqref{eq:normal_eq} admits the closed-form solution:
\begin{equation}
\label{eq:closed_form}
\mathbf{f}^{k+1}
=
\mathbf{v}
+
\frac{
\mathbf{\Phi}^{T}
(\mathbf{g}-\mathbf{\Phi}\mathbf{v})
}{
\mu+\mathbf{\Phi}^{T}\mathbf{\Phi}
}.
\end{equation}
However, with the PSF-inclusive operator $\mathbf{A}$, the matrix $\mathbf{A}^{T}\mathbf{A}$ involves convolutional coupling and is no longer diagonal, making \eqref{eq:closed_form} inapplicable. We address this challenge using conjugate gradient unrolling, as described in Section~\ref{sec:cg_unroll}.

\subsubsection{Dual Variable Update}

The Lagrange multiplier is updated as
\begin{equation}
\label{eq:y_update}
\mathbf{y}^{k+1}
=
\mathbf{y}^{k}
+
\mu^{k+1}
(\mathbf{f}^{k+1}-\mathbf{z}^{k+1}+\mathbf{r}^{k+1}).
\end{equation}

\subsubsection{PSF-Conditioned Penalty Parameter Estimation}

We extend the penalty estimator to receive a global PSF embedding $\mathbf{z}_{\mathrm{psf}}^{\mathrm{g}} \in \mathbb{R}^{64}$ produced by the PSF Encoder. The penalty parameter $\mu^{k+1}$ is computed as
\begin{equation}
\label{eq:mu_est}
\mu^{k+1}
=
\mathrm{Softplus}
\left(
\mathrm{MLP}
\left(
\left[
\mathrm{GAP}(\mathbf{f}^k),\;
\mathbf{z}_{\mathrm{psf}}^{\mathrm{g}}
\right]
\right)
\right)
\cdot
\exp(\beta^k),
\end{equation}
where $[\cdot,\cdot]$ denotes concatenation, $\beta^k$ is a learnable per-stage bias, and GAP denotes global average pooling. Conditioning $\mu$ on the PSF allows the network to adopt stronger regularization when the PSF is severe and lighter regularization when the PSF is closer to a delta function.

\subsection{Conjugate Gradient Unrolling for the f-Subproblem}
\label{sec:cg_unroll}

To solve the normal equation \eqref{eq:normal_eq} under the PSF-inclusive forward model, we employ a $K$-step conjugate gradient (CG) solver that is fully differentiable and supports end-to-end training. The CG solver is warm-started from the ADMM intermediate variable $\mathbf{v}$, which is already a good approximation of the solution from the previous stage.

The CG iteration proceeds as follows. We initialize
\begin{equation}
\label{eq:cg_init}
\mathbf{x}_0 = \mathbf{v}, \quad
\mathbf{r}_0 = \mathbf{b}-\mathbf{Q}(\mathbf{x}_0), \quad
\mathbf{p}_0 = \mathbf{r}_0,
\end{equation}
where
\begin{equation}
\label{eq:Q_def}
\mathbf{Q}(\cdot)=\mathbf{A}^{T}\mathbf{A}(\cdot)+\mu(\cdot).
\end{equation}
For $t=0,1,\ldots,K-1$, the CG update is
\begin{align}
\label{eq:cg_update1}
\alpha_t
&=
\frac{\mathbf{r}_t^{T}\mathbf{r}_t}
{\mathbf{p}_t^{T}\mathbf{Q}(\mathbf{p}_t)}, \\
\mathbf{x}_{t+1}
&=
\mathbf{x}_t + \alpha_t \mathbf{p}_t,
\end{align}
and
\begin{align}
\label{eq:cg_update2}
\mathbf{r}_{t+1}
&=
\mathbf{r}_t - \alpha_t \mathbf{Q}(\mathbf{p}_t), \\
\beta_t
&=
\frac{\mathbf{r}_{t+1}^{T}\mathbf{r}_{t+1}}
{\mathbf{r}_t^{T}\mathbf{r}_t}, \\
\mathbf{p}_{t+1}
&=
\mathbf{r}_{t+1} + \beta_t \mathbf{p}_t .
\end{align}

The operator $\mathbf{Q}$ is never explicitly formed as a matrix. Instead, each application of $\mathbf{Q}(\mathbf{p})$ is computed via one forward pass $\mathbf{A}(\mathbf{p})$ followed by one adjoint pass $\mathbf{A}^{T}(\cdot)$, both implemented using FFT-based convolution. The entire CG unrolling is implemented using standard PyTorch tensor operations, maintaining a complete computational graph for backpropagation.

We set $K=2$ throughout our experiments. The warm start from $\mathbf{v}$ ensures that the initial residual is small, and two CG steps provide sufficient accuracy for the f-update while adding negligible parameters. Specifically, only one learnable scalar log-scale factor is introduced per stage to stabilize training in early epochs.

After the CG solver produces $\mathbf{x}_{\mathrm{cg}}$, we apply a wavelength-adaptive gradient refinement step that exploits the per-wavelength nature of optical PSFs. Let $\mathbf{Z}_{\mathrm{psf}}^{\mathrm{s}} \in \mathbb{R}^{\Lambda \times d}$ denote the per-wavelength PSF embedding produced by the PSF Encoder, where $d$ is the per-wavelength feature dimension. Two lightweight MLPs map $\mathbf{Z}_{\mathrm{psf}}^{\mathrm{s}}$ to per-wavelength step scales and gradient biases:
\begin{align}
\boldsymbol{\sigma}^{k}
&=
1 + \tanh\!\left( \mathrm{MLP}_{\mathrm{scale}}( \mathbf{Z}_{\mathrm{psf}}^{\mathrm{s}} ) \right) \in (0,2)^{\Lambda}, \\
\mathbf{b}^{k}
&=
\tanh\!\left( \mathrm{MLP}_{\mathrm{bias}}( \mathbf{Z}_{\mathrm{psf}}^{\mathrm{s}} ) \right) \in (-1,1)^{\Lambda}.
\end{align}
The refined solution is computed as
\begin{equation}
\label{eq:grad_refine}
\mathbf{f}^{k+1}_{\mathrm{cg}}
=
\mathbf{x}_{\mathrm{cg}}
-
|\eta^{k}| \cdot
\boldsymbol{\sigma}^{k} \odot
\Big(
\mathbf{A}^{T}\!\left( \mathbf{A}(\mathbf{x}_{\mathrm{cg}}) - \mathbf{g} \right)
\;+\;
\mathbf{b}^{k}
\Big),
\end{equation}
where $\eta^{k}$ is a learnable global step size and $\odot$ denotes per-wavelength multiplication broadcasted over the spatial dimensions. This formulation embodies the physical intuition that wavelengths with severe PSF degradation should take more conservative update steps, while wavelengths with milder PSF can take larger steps. The bias $\mathbf{b}^{k}$ further compensates for systematic gradient asymmetries that the finite-step CG cannot fully resolve.

The final f-update is obtained by fusing the CG-refined result with the residual learning pathway through a Fusion Block (FB):
\begin{equation}
\label{eq:f_fusion}
\mathbf{f}^{k+1}
=
\mathrm{FB}
\left(
\mathbf{f}^{k+1}_{\mathrm{cg}},\;
\mathbf{z}^{k+1}-\mathbf{r}^{k+1}
\right).
\end{equation}
The FB module performs purely spatial fusion without PSF conditioning, since PSF information is already injected upstream through the CG solver and gradient refinement steps.

\subsection{Multi-Level PSF Conditioning}
\label{sec:psf_conditioning}

PSF information enters our reconstruction network at three complementary levels, each operating at a distinct stage of the optimization process.

\textit{Level 1: Physical operators.}
The PSF kernels $\{H_c\}$ are directly used in the forward operator $\mathbf{A}$ and the adjoint $\mathbf{A}^{T}$ within the CG solver and the gradient refinement step. The PSF-convolved mask $\mathbf{A}(\mathbf{\Phi})$ is provided to the DPB module in \eqref{eq:r_update}, giving the degradation prior explicit access to the combined mask-PSF degradation pattern. This level ensures that the physical operators faithfully reflect the actual imaging process.

\textit{Level 2: ADMM penalty.}
A learned global PSF embedding $\mathbf{z}_{\mathrm{psf}}^{\mathrm{g}} \in \mathbb{R}^{64}$ conditions the penalty parameter $\mu$ through \eqref{eq:mu_est}, enabling the network to adjust regularization strength based on overall PSF severity.

\textit{Level 3: Wavelength-adaptive gradient.}
A learned per-wavelength PSF embedding $\mathbf{Z}_{\mathrm{psf}}^{\mathrm{s}} \in \mathbb{R}^{\Lambda \times d}$ conditions the gradient refinement step through \eqref{eq:grad_refine}, generating wavelength-specific step sizes and gradient biases that account for spectrally-varying PSF blur.

\begin{figure*}[!t]
\centering
\includegraphics[width=\textwidth]{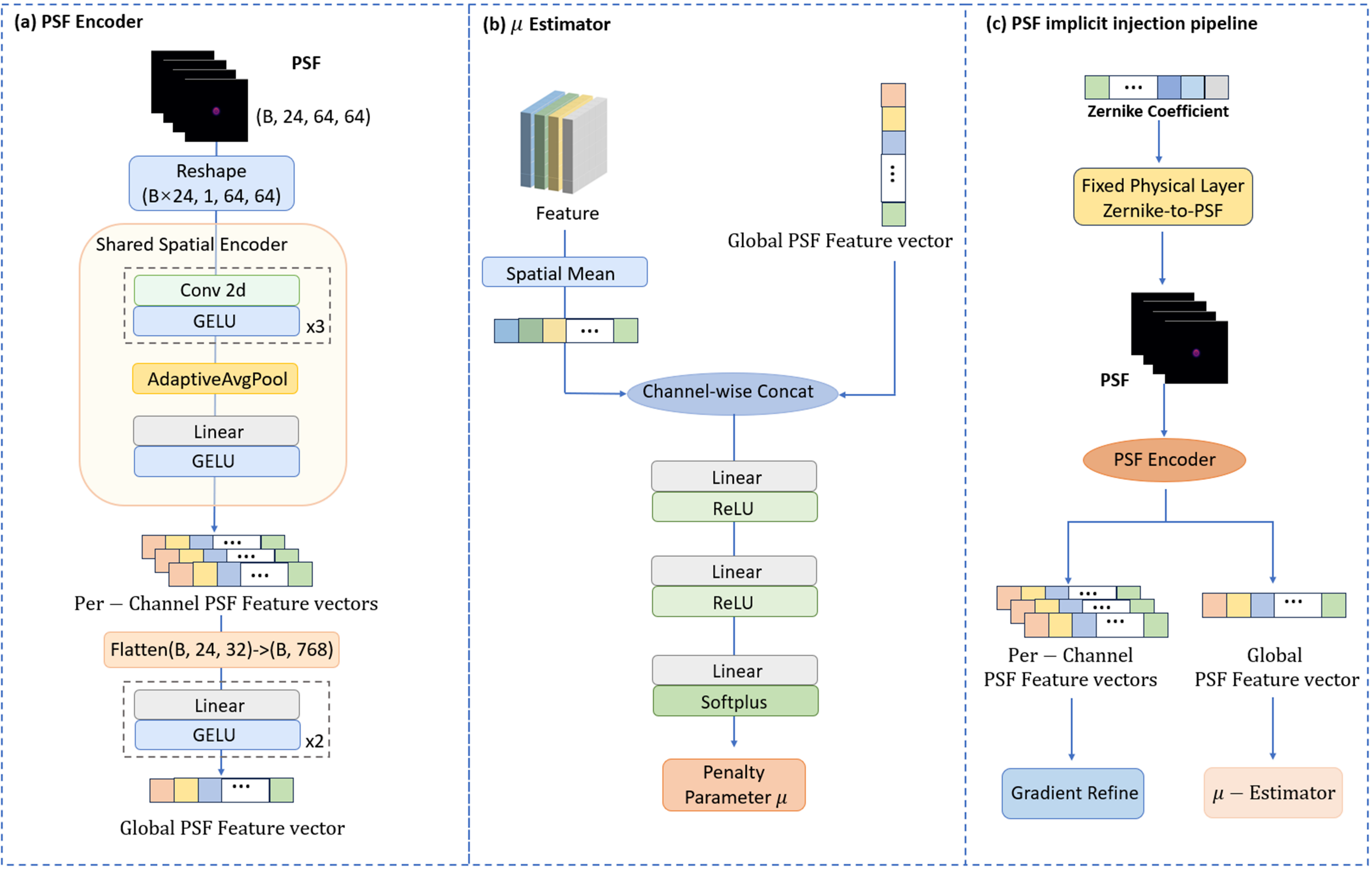}
\caption{PSF conditioning modules. (a) PSF Encoder: each PSF channel is processed independently through a shared spatial CNN to produce a per-channel feature vector $\mathbf{Z}_{\mathrm{psf}}^{\mathrm{s}} \in \mathbb{R}^{B \times 24 \times 32}$. A two-layer MLP further produces the global PSF feature vector $\mathbf{z}_{\mathrm{psf}}^{\mathrm{g}} \in \mathbb{R}^{B \times 64}$. (b) $\mu$ Estimator: the spatial mean of the current reconstruction is concatenated with $\mathbf{z}_{\mathrm{psf}}^{\mathrm{g}}$ and processed through an MLP to produce the ADMM penalty parameter $\mu$. (c) PSF injection pipeline: the global PSF feature vector conditions the $\mu$ estimator, while the per-channel PSF feature vectors condition the gradient refinement module.}
\label{fig:psf_conditioning}
\end{figure*}

\textit{PSF Encoder.}
Both embeddings are produced by a single PSF Encoder, illustrated in Fig.~\ref{fig:psf_conditioning}. The encoder processes each spectral band independently through a shared three-stage CNN, producing a $d$-dimensional feature per wavelength (forming $\mathbf{Z}_{\mathrm{psf}}^{\mathrm{s}}$). The global embedding $\mathbf{z}_{\mathrm{psf}}^{\mathrm{g}}$ is obtained by flattening $\mathbf{Z}_{\mathrm{psf}}^{\mathrm{s}}$ and applying a two-layer MLP. We adopt $d=32$ throughout our experiments.

\subsection{Monte Carlo PSF Training Strategy}
\label{sec:mc_training}

To improve robustness to manufacturing-induced PSF variations, we propose a Monte Carlo (MC) PSF training strategy. During training, each batch is associated with a randomly sampled MC realization from the tolerance analysis:
\begin{enumerate}
\item An MC index $m$ is uniformly sampled from $\{1,\ldots,N_{\mathrm{MC}}\}$.
\item For each sample in the batch, a field index $k$ is uniformly sampled from $\{1,\ldots,N_f\}$.
\item The corresponding PSF $H_{k}^{(m)}$ is retrieved from a pre-rendered cache and used to generate the training measurement according to \eqref{eq:psf_forward}.
\item The same PSF is provided to the network as input.
\end{enumerate}

This strategy exposes the network to the full distribution of plausible PSFs during training, rather than only the nominal design PSF. The pre-rendered MC PSF cache, containing $1000$ realizations, $16$ fields, and $24$ wavelengths, is computed once before training and stored in CPU memory, eliminating the computational overhead of real-time PSF rendering during training.

\subsection{Loss Function}

We adopt a multi-stage loss combining L1 reconstruction loss and spectral angle loss:
\begin{equation}
\label{eq:loss}
\mathcal{L}
=
\frac{1}{S}
\sum_{s=1}^{S}
\frac{s}{S}
\left(
\|\hat{\mathbf{f}}^{s}-\mathbf{f}^{*}\|_1
+
0.1\cdot
\mathcal{L}_{\mathrm{SAM}}
(\hat{\mathbf{f}}^{s},\mathbf{f}^{*})
\right),
\end{equation}
where $S$ is the number of unfolding stages, $\hat{\mathbf{f}}^{s}$ is the output of stage $s$, $\mathbf{f}^{*}$ is the ground truth, and $\mathcal{L}_{\mathrm{SAM}}$ is the spectral angle mapper loss. The linearly increasing weight $s/S$ emphasizes later stages, which produce more refined reconstructions. The spectral angle loss encourages spectral fidelity in addition to spatial accuracy.


\section{Experiments}

\subsection{Experimental Setup}

\subsubsection{Datasets}
We evaluate our method on the widely used CAVE~\cite{CAVE2007} and KAIST~\cite{KAIST2017} hyperspectral image datasets. Following the standard protocol in~\cite{MST2022,DAUHST2022,DPU2024,CosineAnnealing2016}, 32 scenes from the CAVE dataset (each with 31 spectral bands) are used for training, and 10 scenes from the KAIST dataset are used for testing. We select 24 spectral bands spanning 470--700~nm to match the spectral range of our DD-CASSI optical design. Training patches of size $128\times128\times24$ are randomly cropped with a core region of $64\times64$, and the outer margin serves as a boundary buffer to mitigate edge artifacts.

\subsubsection{PSF Configuration}
The field-dependent PSF is derived from a DD-CASSI optical design in Zemax OpticStudio with $f/10.1$, pixel size $3.45\;\mu\text{m}$. The nominal Zernike coefficients are exported for a $4\times4$ grid of field positions (16 fields) across 24 wavelengths, yielding 12 Zernike terms per field-wavelength combination (terms 4--15, excluding piston, tip, and tilt). For Monte Carlo tolerance analysis, 1000 realizations are generated by perturbing the optical surface parameters according to manufacturing specifications. All PSFs are pre-rendered at $128\times128$ resolution and cached before training.

\subsubsection{Evaluation Protocol}
For testing, each $256\times256$ KAIST scene is divided into $4\times4=16$ overlapping blocks of $128\times128$, each corresponding to one field position. Each block is independently degraded by its field-specific PSF, modulated by the coded aperture mask, compressed, and corrupted with Gaussian noise ($\sigma=0.005$). The network reconstructs each block independently, and the full scene is assembled using overlap-and-average with an $80\times80$ output window per block. We report three standard metrics: peak signal-to-noise ratio (PSNR), structural similarity index (SSIM), and spectral angle mapper (SAM).

\subsubsection{Testing Conditions}
We evaluate under two PSF conditions:
\begin{itemize}
\item \textit{Nominal test:} Measurements are generated using the nominal (design) PSF, and the same nominal PSF is provided to the network.
\item \textit{MC-matched test:} Measurements are generated using a Monte Carlo sampled PSF, and the same MC PSF is provided to the network. Results are averaged over 5 MC realizations.
\end{itemize}
The MC-matched test simulates a tolerance-aware evaluation condition 
where the network receives the same PSF realization that generated 
the measurement, corresponding to the scenario in which the PSF is 
calibrated after manufacturing.

\subsubsection{Implementation Details}
Our network is implemented in PyTorch and trained on a single NVIDIA RTX 4090 GPU. We use the Adam optimizer with $\beta_1=0.9$, $\beta_2=0.999$, initial learning rate $2\times10^{-4}$, and cosine annealing schedule~\cite{CosineAnnealing2016} over 300 epochs with minimum learning rate $10^{-6}$. The batch size is 16. The number of unfolding stages is $S=5$, the CG step count is $K=2$, and the PSF embedding dimension is 64. Training noise is sampled log-uniformly from $[10^{-3},10^{-1.5}]$ per batch to improve noise robustness. Gradient clipping with max norm 2.0 is applied.

\subsubsection{Comparison Methods}
We compare our PSF-aware method (denoted as \textbf{Ours}) against the following PSF-agnostic baselines:
\begin{itemize}
\item \textbf{B (DPU baseline):} The original DPU~\cite{DPU2024} with 5 stages and 24 channels (${\sim}1.27$M parameters), using the conventional closed-form f-update.
\item \textbf{B+ (DPU enlarged):} A capacity-enlarged variant of DPU with widened internal channels (32 channels, ${\sim}2.12$M parameters) to control for the effect of increased model capacity.
\item \textbf{MST}~\cite{MST2022}: A spectral-wise Transformer baseline.
\item \textbf{DAUHST}~\cite{DAUHST2022}: A degradation-aware unfolding method with half-shuffle attention.
\item \textbf{RDLUF}~\cite{RDLUF2023}: A residual degradation learning unfolding framework.
\end{itemize}
All comparison methods are trained under identical conditions (MC PSF data augmentation, same training schedule) but do not receive PSF as input---the PSF is used only for generating training measurements.

\subsection{Main Results}

Table~\ref{tab:main} presents the quantitative comparison between our method and existing reconstruction methods under both testing conditions. All comparison methods are retrained from scratch under our DD-CASSI imaging model with field-dependent PSF degradation, using identical training data and schedule. These methods do not receive PSF information---the PSF is used only for generating training measurements.

\begin{table*}[!t]
\caption{Quantitative Comparison on 10 KAIST Scenes Under Field-Dependent PSF Degradation}
\label{tab:main}
\centering
\small
\setlength{\tabcolsep}{5pt}
\begin{tabular}{llccccccc}
\toprule
\multirow{2}{*}{Method} & \multirow{2}{*}{Type} & \multirow{2}{*}{Params} & \multicolumn{3}{c}{Nominal Test} & \multicolumn{3}{c}{MC-matched Test} \\
\cmidrule(lr){4-6} \cmidrule(lr){7-9}
& & & PSNR$\uparrow$ & SSIM$\uparrow$ & SAM$\downarrow$ & PSNR$\uparrow$ & SSIM$\uparrow$ & SAM$\downarrow$ \\
\midrule
DAUHST~\cite{DAUHST2022} & Unfolding & 1.78M & 27.52 & 0.8675 & 17.74 & 27.21 & 0.8624 & 17.95 \\
RDLUF~\cite{RDLUF2023} & Unfolding & 2.04M & 27.39 & 0.8649 & 18.43 & 27.08 & 0.8605 & 18.60 \\
MST~\cite{MST2022} & Transformer & 3.66M & 28.14 & 0.8824 & 16.23 & 27.93 & 0.8781 & 16.36 \\
DPU-B~\cite{DPU2024} & Unfolding & 1.27M & 27.80 & 0.8752 & 17.34 & 27.59 & 0.8701 & 17.55 \\
DPU-B+~\cite{DPU2024} & Unfolding & 2.12M & 28.83 & 0.8931 & 15.86 & 28.60 & 0.8882 & 16.07 \\
\midrule
\textbf{Ours} & \textbf{Unfolding} & \textbf{1.42M} & \textbf{30.53} & \textbf{0.9117} & \textbf{14.38} & \textbf{30.43} & \textbf{0.9091} & \textbf{14.42} \\
\bottomrule
\end{tabular}
\end{table*}

Our method outperforms all comparison methods under the PSF degradation setting. Compared to the best-performing PSF-agnostic baseline DPU-B+, our method achieves +1.70~dB improvement in PSNR under the nominal test and +1.83~dB under the MC-matched test, while using 33\% fewer parameters (1.42M vs.~2.12M). The improvement over other state-of-the-art methods is even more pronounced: +3.01~dB over DAUHST, +3.14~dB over RDLUF, and +2.39~dB over MST. These results demonstrate that existing reconstruction methods, which assume ideal optics, suffer significant performance degradation when optical PSF is present, while our PSF-aware approach effectively compensates for this degradation with a more parameter-efficient design.

Notably, our method exhibits superior robustness under PSF perturbation. The PSNR drop from nominal to MC-matched conditions is only 0.10~dB for our method, compared to 0.21--0.31~dB for the baselines. This indicates that incorporating PSF into both the forward operator and the optimization process makes the reconstruction inherently robust to PSF variations.

\subsection{Per-Field Analysis}

\begin{figure*}[!t]
\centering
\includegraphics[width=\textwidth]{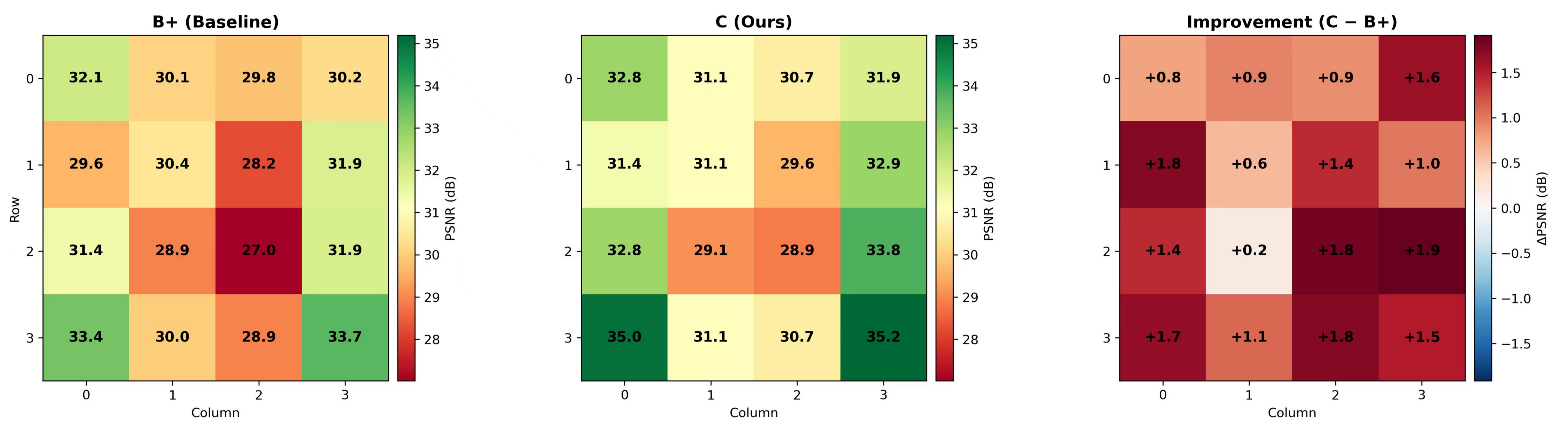}
\caption{Per-field PSNR heatmap ($4\times4$ field grid, averaged over 10 scenes). Left: DPU-B+. Middle: Ours. Right: Improvement (Ours $-$ B+). Our method achieves higher PSNR at all 16 field positions.}
\label{fig:heatmap}
\end{figure*}

A key advantage of our PSF-aware method is improved reconstruction uniformity across the field of view. Fig.~\ref{fig:heatmap} shows the per-field PSNR heatmap for DPU-B+ and our method, averaged over all 10 test scenes.

The improvement heatmap shows that our method achieves higher PSNR at all 16 field positions, with per-field gains ranging from $+0.2$~dB to $+1.9$~dB (average $+1.3$~dB). The improvement is spatially varying, reflecting the field-dependent nature of PSF-induced degradation.

\begin{table*}[!t]
\caption{Ablation Study on KAIST Test Set Under Nominal and MC PSF Conditions}
\label{tab:ablation}
\centering
\small
\setlength{\tabcolsep}{5pt}
\begin{tabular}{lcccccc}
\toprule
\multirow{2}{*}{Configuration} & \multirow{2}{*}{Params} & \multicolumn{2}{c}{Nominal Test} & \multicolumn{2}{c}{MC-matched Test} & \multirow{2}{*}{$\Delta$PSNR (Nom./MC)} \\
\cmidrule(lr){3-4} \cmidrule(lr){5-6}
& & PSNR$\uparrow$ & SSIM$\uparrow$ & PSNR$\uparrow$ & SSIM$\uparrow$ & \\
\midrule
DPU-B (baseline)           & 1.27M & 27.80 & 0.8752 & 27.59 & 0.8701 & --- / --- \\
DPU-B+ (capacity-matched)  & 2.12M & 28.83 & 0.8931 & 28.60 & 0.8882 & +1.03 / +1.01 \\
\midrule
w/o PSF Encoder            & 1.30M & 30.39 & 0.9092 & 30.26 & 0.9066 & +2.59 / +2.67 \\
w/o PSF in $\mu$           & 1.42M & 30.32 & 0.9094 & 30.16 & 0.9063 & +2.52 / +2.57 \\
w/o Wavelength-Adaptive GR & 1.42M & 30.36 & 0.9047 & 30.20 & 0.9016 & +2.56 / +2.61 \\
w/o CG (closed-form)       & 1.42M & 29.57 & 0.8972 & 29.41 & 0.8933 & +1.77 / +1.82 \\
w/o GradStep (CG only)     & 1.42M & 29.86 & 0.9044 & 29.79 & 0.9018 & +2.06 / +2.20 \\
\midrule
\textbf{Full model (Ours)} & \textbf{1.42M} & \textbf{30.53} & \textbf{0.9117} & \textbf{30.43} & \textbf{0.9091} & \textbf{+2.73 / +2.84} \\
\bottomrule
\end{tabular}
\end{table*}

\subsection{Ablation Study}
\textit{CG Unrolling (+0.96~dB).}
Replacing the CG solver with a closed-form update (which assumes $\mathbf{A}^{T}\mathbf{A}$ is diagonal, equivalent to ignoring the PSF) yields a 0.96~dB drop in PSNR. This is the largest single-component contribution and confirms that the non-diagonal structure of $\mathbf{A}^{T}\mathbf{A}$ under the PSF-inclusive model requires an iterative solver.

\textit{Gradient Refinement Module (+0.67~dB).}
Bypassing the entire gradient refinement step (using $\mathbf{x}_{\mathrm{cg}}$ directly without further correction) yields a 0.67~dB drop. This confirms that the finite-step CG solver leaves residual errors that benefit from learned correction.

\textit{Wavelength-Adaptive Step (+0.17~dB).}
Disabling the wavelength-adaptive components within the gradient refinement (replacing $\boldsymbol{\sigma}^k$ with a uniform global step and setting $\mathbf{b}^k=\mathbf{0}$) yields a 0.17~dB drop. While smaller in absolute magnitude than CG unrolling, this contribution validates the physical intuition that spectrally-varying PSF blur warrants per-wavelength update rates.

\textit{PSF-Conditioned Penalty (+0.21~dB).}
Disabling PSF conditioning of $\mu$ (zeroing out the PSF input to the penalty estimator) yields a 0.21~dB drop, confirming that the global PSF embedding provides useful information for adaptive regularization.

\textit{PSF Encoder Total Contribution (+0.14~dB).}
Disabling the PSF encoder entirely (zeroing both global and per-wavelength embeddings) yields a 0.14~dB drop. Interestingly, this is smaller than the sum of the individual pathway contributions (+0.21 from $\mu$ and +0.17 from wavelength step, totaling +0.38~dB). This apparent inconsistency arises because, when the PSF embeddings are zero, the bias terms in the wavelength-scale and gradient-bias MLPs can still learn a fixed (PSF-independent) schedule, partially compensating for the missing input. This observation highlights that the value of PSF conditioning lies in providing \emph{adaptive}, sample-specific schedules rather than a fixed one.

\textit{Total method contribution.}
Compared with the DPU-B baseline, our full method improves PSNR by +2.73~dB on the nominal test and +2.84~dB on the MC-matched test. Note that the sum of individual component contributions (+2.16~dB) is smaller than the total (+2.73~dB), indicating non-trivial positive interactions among the components.

\subsection{CG Steps Ablation}

\begin{figure*}[!t]
\centering
\includegraphics[width=\textwidth]{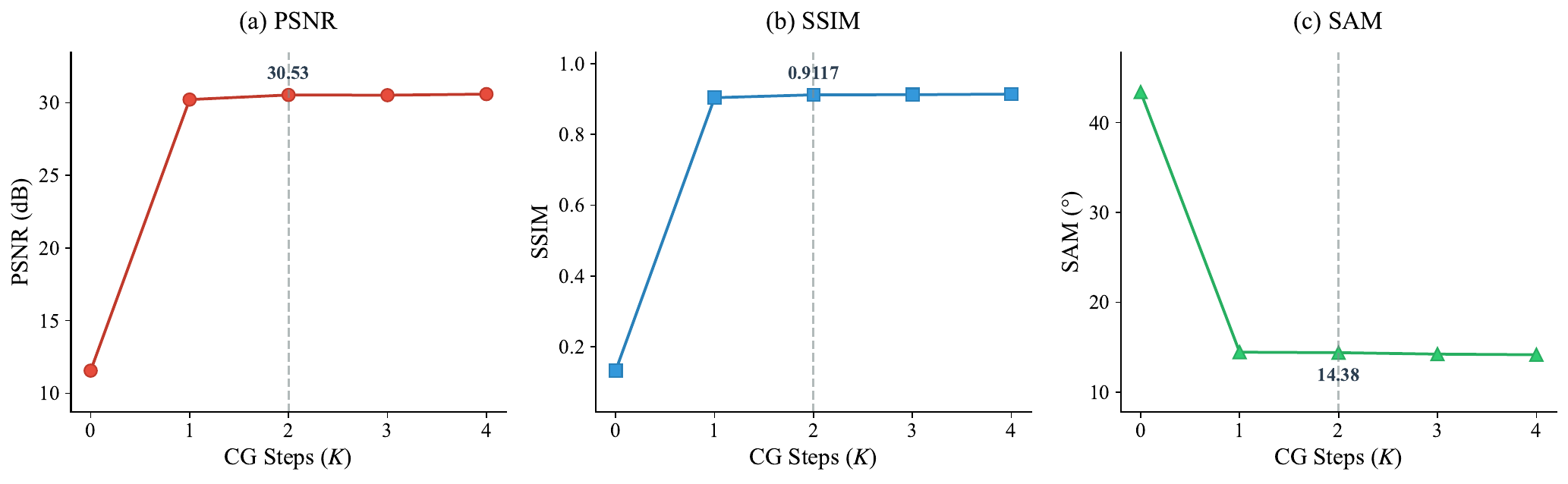}
\caption{Effect of CG steps $K$ on reconstruction quality. $K=0$ (no CG) yields only 11.56~dB, confirming the necessity of the CG solver. $K=2$ (used during training, marked with square markers) achieves near-optimal performance and is adopted as the default. Increasing $K$ to 4 only marginally improves PSNR ($\leq$+0.06~dB) at the cost of approximately twice the CG-related computational cost.}
\label{fig:cg_ablation}
\end{figure*}

To validate the choice of $K=2$ CG steps, we evaluate the trained full model with different CG step counts at test time. Fig.~\ref{fig:cg_ablation} shows the results.

$K=0$, which corresponds to using the warm-start point directly without any CG iteration, drops PSNR to 11.56 dB, confirming that CG iterations are essential for the PSF-inclusive forward model. $K=1$ substantially recovers performance (30.22 dB) but still falls 0.31 dB short of $K=2$. Increasing $K$ to 3 or 4 yields only marginal improvements ($\leq +0.06$ dB) at the cost of doubled FFT operations. We therefore adopt $K=2$ as the default setting, balancing accuracy and computational efficiency.

\subsection{Qualitative Results}

\begin{figure*}[!t]
\centering
\includegraphics[width=\textwidth]{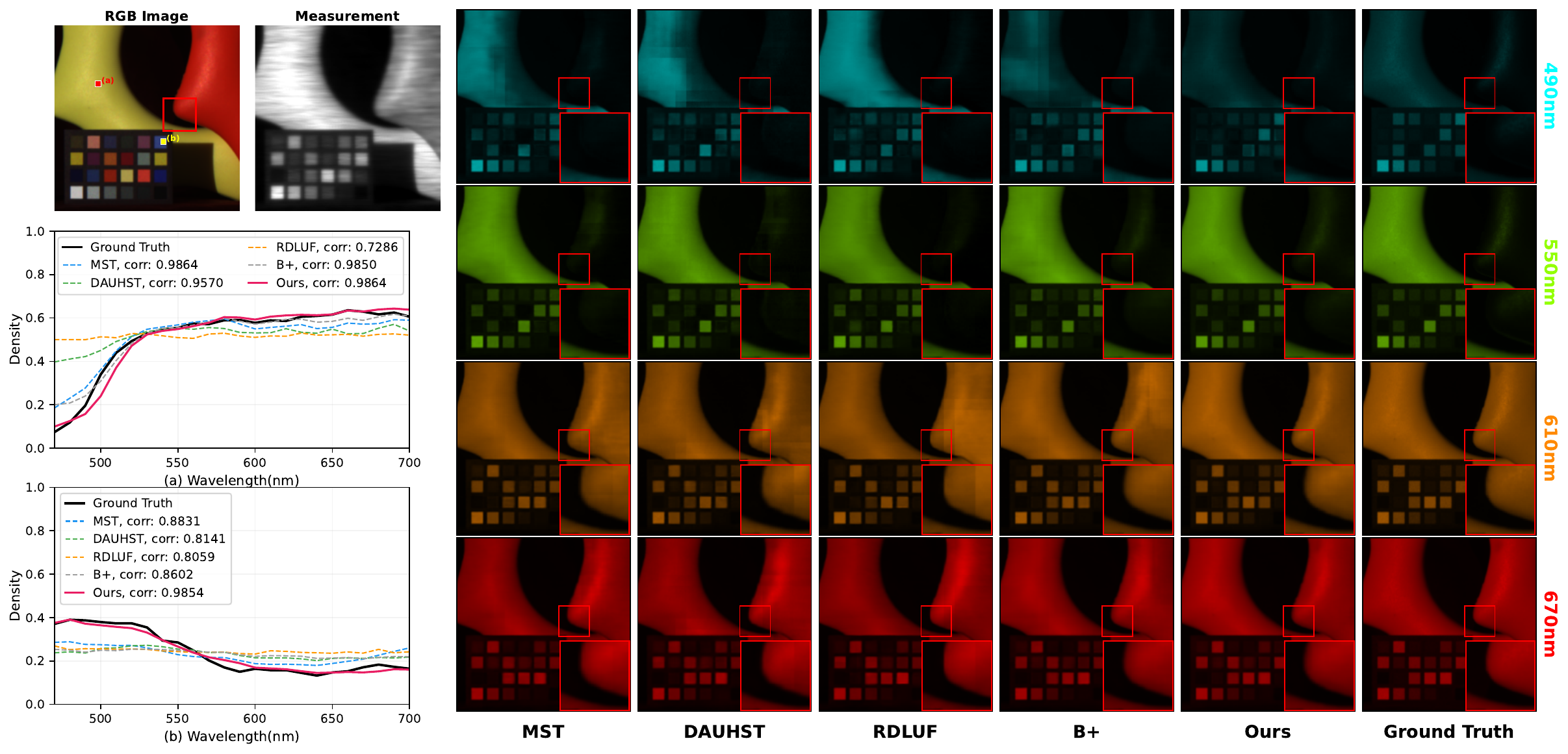}
\caption{Qualitative comparison on a representative KAIST test scene. Left: pseudo-RGB image, compressed measurement, and spectral density curves at two selected pixels (marked as (a) and (b)). Right: reconstructed images at four wavelengths with zoom-in regions (red boxes). Our method produces sharper textures, fewer artifacts, and higher spectral correlation than all comparison methods.}
\label{fig:comparison}
\end{figure*}

Fig.~\ref{fig:comparison} presents the qualitative comparison on a representative test scene. The left panel shows the pseudo-RGB image, compressed measurement, and spectral curves at two selected pixel locations. The right panel displays the reconstructed images at four wavelengths (490, 550, 610, and 670~nm) for all comparison methods.

Our method produces spectral curves that closely match the ground truth across the entire 470--700~nm range, achieving the highest spectral correlation coefficients among all methods. The reconstructed images show that PSF-agnostic methods (MST, DAUHST, RDLUF, DPU-B+) produce blurred structures and spectral artifacts, particularly in the zoom-in regions where PSF degradation is most pronounced. In contrast, our method recovers sharper edges, finer textures, and more accurate spectral content, demonstrating the effectiveness of explicit PSF modeling in the reconstruction pipeline.

\subsection{Verification of PSF Utilization}

A natural question is whether the observed improvement stems from
genuine PSF-aware reconstruction or merely from the additional
network capacity introduced by the PSF encoder and PSF-conditioned modules.

To answer this, we conduct a \emph{PSF-mismatch} experiment:
measurements are generated using an MC-sampled PSF, but the
network receives only the nominal PSF.

Under this deliberate mismatch, our method's PSNR drops
progressively with calibration error (from 30.53\,dB matched to
28.27\,dB and 26.66\,dB at the first and second MC quartiles),
falling below the PSF-agnostic baseline B+ beyond the 25th
percentile. This result provides two important confirmations:

\begin{enumerate}
\item \textbf{Genuine PSF utilization:} The network actively
incorporates the provided PSF into its reconstruction process.
If the PSF pathway were merely adding capacity without
functional use, mismatched PSF input would not cause degradation.
\item \textbf{Consistency with non-blind inverse theory:} Any
non-blind method that faithfully inverts a provided forward model
will produce biased solutions when the model is incorrect. This
behavior is shared by classical Wiener filtering, Richardson-Lucy
deconvolution, and all model-based iterative methods---it reflects
correct algorithmic behavior rather than a method-specific limitation.
\end{enumerate}

Field-dependent PSFs can be obtained through several complementary 
routes: (i) direct optical testing such as wavefront sensing or 
point-source calibration; (ii) coded-aperture-based PSF estimation, 
in which the calibration images---acquired by illuminating the coded 
aperture with monochromatic light---are treated as blurred versions 
of the mask and the PSF is recovered via regularized least-squares, 
as proposed by Song~et~al.~\cite{Song2022}; and (iii) physical-optics 
computation from the optical design, in which the Huygens PSF is 
evaluated from ray-traced wavefronts at the exit pupil, as adopted by 
Yue~et~al.~\cite{Yue2024} and by the present work. These options span 
a spectrum from post-manufacturing measurement to pre-manufacturing 
simulation, indicating that the PSF input required by our framework 
is practically obtainable in typical CASSI deployments. The MC-matched 
results in Table~\ref{tab:main} (30.43~dB) demonstrate that when the 
provided PSF is consistent with the degradation model, our method 
delivers substantial gains over PSF-agnostic approaches.

\subsection{Computational Complexity}

Table~\ref{tab:complexity} summarizes the computational complexity of the compared methods.

Our method uses 1.42M parameters, comparable to the most compact baseline DPU-B (1.27M) and 33\% fewer than DPU-B+ (2.12M). In terms of FLOPs, our method (5.65G) is nearly identical to DPU-B (5.58G), confirming that the CG solver and PSF-conditioned modules introduce minimal computational overhead. The slightly higher inference time (0.0503s vs.\ 0.0416s for DPU-B) is attributable to the $2K{=}4$ FFT-based forward/adjoint passes per stage required by the CG solver. Overall, the absolute inference time remains modest (50ms per 128×128 block), comparable to other unfolding methods.

\begin{table}[!t]
\caption{Computational Complexity Comparison}
\label{tab:complexity}
\centering
\begin{tabular}{l ccc}
\toprule
Method & Params (M) & FLOPs (G) & Inference (s) \\
\midrule
MST         & 3.66 & \textbf{3.98} & \textbf{0.0109} \\
DAUHST-5stg & 1.78 & 6.10 & 0.0207 \\
RDLUF-5stg  & 2.04 & 14.72 & 0.0455 \\
DPU-B+      & 2.12 & 9.44 & 0.0444 \\
DPU-B       & \textbf{1.27} & 5.58 & 0.0416 \\
\textbf{Ours} & 1.42 & 5.65 & 0.0503 \\
\bottomrule
\end{tabular}
\end{table}

\section{Conclusion}

In this paper, we have presented a PSF-aware deep unfolding network for coded aperture snapshot spectral imaging that explicitly incorporates field-dependent, wavelength-dependent optical point spread functions into the reconstruction framework. Our approach addresses the algorithmic challenge that arises when the PSF is incorporated into the forward model: the diagonal structure of the normal equation $\mathbf{\Phi}^T\mathbf{\Phi}$ is broken, rendering the closed-form data-fidelity updates used in existing unfolding methods inapplicable.

The proposed method introduces three key contributions. First, we formulate a PSF-inclusive forward model that replaces the conventional mask-only sensing operator. Since this renders the normal equation non-diagonal, we develop a $K$-step conjugate gradient unrolling scheme that provides a principled solution while maintaining differentiability. Second, we propose a wavelength-adaptive gradient refinement module that uses a learned per-wavelength PSF embedding to generate spectral-channel-specific step sizes and gradient biases. Third, we propose a PSF-conditioned penalty estimator and a Monte Carlo PSF training strategy to handle manufacturing-induced variations. Importantly, all PSF conditioning operates in the optimization space, directly affecting the CG iterates, penalty parameter, and gradient step.

Experimental results on the CAVE and KAIST datasets demonstrate that our method achieves 30.53~dB PSNR, +1.70~dB over the larger PSF-agnostic baseline, while using only 1.42M parameters---33\% fewer than the baseline and 61\% fewer than the best Transformer competitor. Comprehensive ablation studies confirm that each proposed component contributes meaningfully, with the CG unrolling providing the largest single-component gain (+0.96~dB). Our method also exhibits superior robustness to PSF perturbation, with only 0.10~dB degradation under Monte Carlo conditions compared to 0.21--0.31~dB for baselines.

Our work opens several directions for future research. First, the current method assumes that the PSF is accurately known; developing robustness to PSF calibration errors through uncertainty-aware conditioning would broaden practical applicability. Second, extending the framework to handle spatially-continuous PSF variation, rather than a discrete field grid, through interpolation or implicit neural representations could improve the spatial resolution of the PSF model. Third, joint optimization of the optical design and the reconstruction network in an end-to-end fashion could lead to CASSI systems that are co-designed for optimal reconstruction quality. 

Finally, although hardware measurements are not included in the present study, 
the proposed framework is evaluated under a complete optical-design 
model with field-, wavelength-, and tolerance-dependent perturbations. 
Validation on real CASSI hardware measurements remains an important 
direction for future work and would provide the ultimate test of the 
proposed approach.


\end{document}